\documentclass{jfm}
\usepackage{graphicx}

\usepackage{newtxtext}
\usepackage{newtxmath}
\usepackage{natbib}
\usepackage{hyperref}
\usepackage{svg}
\hypersetup{
    colorlinks = true,
    urlcolor   = blue,
    citecolor  = black,
}

\usepackage{xcolor}            
\usepackage[normalem]{ulem}

\makeatletter
\gdef\@underjournal{}

\def\footerflagdefns#1{}
\def\absfooterflag{}
\def\pagelimitfooter{}

\def\ps@titlepage{\leftskip\z@\let\@mkboth\@gobbletwo\vfuzz=5\p@
  \def\@oddhead{\vbox{\vspace*{-1pc}\hbox to \textwidth{\@j@urnal \hfil\llap{\thepage}}\par\vskip4pt}}
  \def\@evenhead{\@j@urnal \hfil\llap{\thepage}}%
  \def\@oddfoot{}%
  \def\@evenfoot{}%
  \def\sectionmark##1{}%
  \def\subsectionmark##1{}%
}

\def\ps@spheadings{\let\@mkboth\markboth
  \def\@oddhead{\hfil{\itshape\@shorttitle}\hfil\llap{\thepage}}%
  \def\@evenhead{\rlap{\thepage}\hfil\itshape\@shortauthor\hfil}%
  \def\@oddfoot{}%
  \def\@evenfoot{}%
  \def\sectionmark##1{\markboth{##1}{}}%
  \def\subsectionmark##1{\markright{##1}}%
}
\makeatother

\title{Wave Attenuation in Drifting Sea Ice: A Mechanistic Model for Observed Decay Profiles}

\author{Rhys Ransome\aff{1}
  \corresp{\email{R.Ransome@uea.ac.uk}},
  Davide Proment\aff{1}, Ian A. Renfrew\aff{2}
 \and Alberto Alberello\aff{1}}

\affiliation{%
  \aff{1}School of Engineering, Mathematics and Physics, University of East Anglia, Norwich NR4 7TJ, UK\\
  \aff{2}School of Environmental Sciences, University of East Anglia, Norwich NR4 7TJ, UK
}

\begin{document}
\maketitle

\begin{abstract}
Wave-sea ice interactions shape the transition zone between open ocean and pack ice in the polar regions.
Most theoretical paradigms, implemented in coupled wave-sea ice models, predict exponential decay of the wave energy but some recent observations deviate from this behaviour.
Expanding on a framework based on wave energy dissipation due to ice-water drag, we account for drifting sea ice to derive an improved model for wave energy attenuation. 
Analytical solutions replicate the observed non-exponential wave energy decay and the spatial evolution of the effective attenuation rate in Antarctic sea ice.
\end{abstract}


\section{Introduction}

Sea ice that seasonally forms in the polar regions regulates heat and momentum fluxes between the ocean and the atmosphere over large scales, thereby playing a crucial role in the global weather and climate system
\citep[e.g.][]{IPCC_2021_WGI,bennetts2024rev,selivanova2024past}.
The transition zone between the open ocean and the pack ice, often referred to as the marginal ice zone \citep[MIZ;][]{wadhams1986}, is the most dynamic, yet poorly understood, component of such system \citep{landwwehr2021esd_short,vichi2022tc,day2024jgr}.

The interaction of ocean waves propagating into the sea ice is a defining feature of the Arctic and Antarctic MIZ \citep{kohout2014nature,stopa2018pnas,alberello2022nc,brouwer2022altimetric,day2024jgr}, and in the Southern Ocean this region extends for hundreds of kilometres. 
Wave energy decays with distance into the MIZ, with the attenuation rate $\alpha$, i.e. $E(x) \propto \exp(-\alpha x)$, assumed to be proportional to the wave frequency $f$ to some power \citep[$\alpha\propto f^p$; e.g.][]{meylan2018jgr}.
The exact frequency dependency varies according to the dominant attenuation mechanism, and various options are implemented in operational third generation spectral wave models \citep[e.g. WaveWatchIII;][]{montiel2025jgr} that are being coupled into global- and hemispheric-scale ocean-sea ice models.
Most models predict a constant attenuation rate in homogeneous sea ice conditions and, consequently, exponential wave energy decay.

Overall, measurements conform reasonably well to the exponential attenuation paradigm and agree on the magnitude of the attenuation rate $\alpha$, within the range \citep[$10^{-4}$\,m$^{-1}<\alpha<10^{-5}$\,m$^{-1}$;][]{kohout2020ag,montiel2022jpo}.
However, recent satellite measurements of wave height in the Antarctic MIZ across seasons revealed an almost linear increase of the average attenuation rate with distance from the sea ice edge \citep[up to $\approx 10\times$;][]{voermans2025finely} with individual transects showing sharp peaks in attenuation rate. 
Possible causes of deviation from the exponential profile were attributed to the misalignment between the transect and wave direction, and inhomogeneous sea ice conditions (sea ice concentration and thickness).
Lacking a firm theoretical foundation, \citet{voermans2025finely} propose an empirical wave attenuation formulation as an interim solution for implementation in wave models,
i.e. $\alpha = \beta x/x_{\text{MIZ}}-\gamma$ 
in which $\beta$ and $\gamma$ are empirical fitting parameters and 
$x_{\text{MIZ}}$ is the extent of the MIZ. The formulation leads to exponential energy decay at a rate $\beta x^2$.

Amongst the various theoretical wave attenuation models, the ones that attribute wave energy dissipation to quadratic ice-water drag deviate from exponential decay in form.
Ice-water drag 
{models capture the bulk of the} 
energy loss when sea ice comprises floes with small diameter, relative to wavelength, in high concentration 
{\citep{shen1998wave,kohout2011wave,herman2019wave}}  as further supported by experimental observations \citep{smith2020pancake}. This makes these models well suited to the Antarctic MIZ.
The attenuation rate due to drag is proportional to the square of the relative velocity between fluid orbital motion and sea ice \citep{herman2019wave}.
Assuming a stationary semi-infinite sea ice domain, analytical, non-exponential, solutions for the amplitude profile have been obtained under slightly different approaches by \citet{kohout2011wave}, for the amplitude envelope, and \citet{herman2019wave}, for the phase-averaged amplitude.
The former is a lower bound for the amplitude, i.e. a higher bound for the attenuation rate.

The drift of sea ice is a feature of MIZ dynamics throughout the year \citep{kwok2017sea,day2024jgr}.
The Southern Ocean, under persistent waves \citep{derkani2020wind}, is characterised by unconsolidated small floes ($\approx10$\,m in diameter) in high concentration \citep[$>80\%$;][]{alberello2019tc,day2024jgr} with little internal stresses \citep{alberello2020jgr,womack2022jog}. In these conditions, wind is the main driver of sea ice drift \citep{holland2012wind,kwok2017sea}.
Through air-ice drag coefficients of order $10^{-3}$ \citep{wamser1993drag,elvidge2016observations}, sea ice drifts at 0.1--0.4\,m/s with low variability over large scales \citep[10--100\,km;][]{holland2012wind}.
Nevertheless, the drift of the MIZ is particularly intense during extreme events when concurrent energetic ocean wave are able to penetrate hundreds of kilometres into the sea ice, as demonstrated by recent observations \citep{vichi2019grl,alberello2020jgr,womack2022jog,womack2024contrast}. These events are likely the main contributors to the overall MIZ dynamics.
However, to the best of our knowledge, the interplay between drift and wave attenuation has not been explored before. 

Several of the world’s leading meteorological agencies have moved to using coupled atmosphere–ocean–sea-ice forecasting systems in the last few years to improve forecasting of near-surface temperature, humidity, winds and low-level clouds, as well as sea-ice distribution in the polar regions. In this regard, prediction in the MIZ is proving a particularly difficult challenge \citep{day2022benefits,barrell2025rapid}. One aspect of focussed development is the coupling of physics-based ocean wave and sea-ice models, within a dynamical ocean model, so subject to drift \citep{kousal2025bams}, making this study timely with the potential to inform this development.

 Here, we develop a new analytical framework for the attenuation of monochromatic waves due to drag, explicitly accounting for constant sea ice drift.

We extend the previous works of \citet{kohout2011wave} and \citet{herman2019wave} where drift was neglected on the grounds that, in typical  oceanic conditions, drift velocities are two orders of magnitude smaller than the wave group velocity.

Despite its apparently negligible effect, we reveal distinctive consequences of including such drift, most notably the emergence of a location into the ice field where the wave amplitude vanishes and the attenuation rate becomes unbounded.

We further demonstrate the capability of this enhanced model to reproduce recent observations of wave attenuation in the Antarctic MIZ.

\section{Mathematical Formulation and Analytical Solutions}
\label{sec:headings}

The one-dimensional wave energy transport equation for energy density per unit surface area, $E(x,t)$, in sea ice-covered regions and excluding all other source terms is:
\begin{equation}\label{energyequatione}
\frac{\partial E}{\partial t} + c_g\frac{\partial E}{\partial x}= S_{\text{ice }}=  S_{\text{sd}}+S_{\text{exp}},
\end{equation}
where $c_g $ denotes the group velocity

and,
{similarly} to \citet{kohout2011wave}, the effect of the ice $(S_{\text{ice}}$) is explicitly expressed as a linear superposition of all the mechanisms {with constant decay,} that lead to exponential attenuation ($S_{\text{exp}}$), and the ones due to ocean-ice skin drag $(S_{\text{sd}}$). 
Note that the assumption of exponential decay for all the mechanisms other than drag will facilitate analytic handling.
Here, in agreement with most wave-in-ice parametrisations in WaveWatchIII \citep{montiel2025jgr}, we assume the open water dispersion relation between angular frequency $\omega$ and wave-number $k$ to hold in sea ice.

The transport equation in terms of wave amplitude $a(x,t)$, using $E= \frac{1}{2} \rho g a^2$ where $\rho$ is water density and $g$ acceleration due to gravity, becomes:
\begin{equation}\label{energyequationa}
a\frac{\partial a}{\partial t} + ac_g\frac{\partial a}{\partial x}= s_{\text{ice }}=  s_{\text{exp}} + s_{\text{sd}},
\end{equation}
where $s=S/\rho g$. In agreement with previous  work  \citep[e.g.][]{kohout2011wave,herman2019wave}, in ice covered regions we define the exponential source term as
\begin{equation}\label{exp}
 s_{\text{exp}} =-c_g \alpha_\textrm{exp} a^2,
\end{equation}
where $\alpha
_\textrm{exp}$ is the constant amplitude attenuation rate for all other contributing non skin-drag mechanisms e.g. wave scattering \citep{wadhams1978wave}, viscous dissipation \citep{weber1987wave,keller1998gravity} and visco-elasticity \citep{robinson1990modal,fox1994oblique,wang2010gravity}. Its inclusion provides flexibility for future operational model implementations.

In ice covered regions, for a monochromatic wave, the skin drag stress at the ice-ocean interface follows a standard quadratic law, $\tau = \rho C_d|u_{\text{orb}}-v|(u_{\text{orb}}-v)$ \citep{shen1998wave}, with $C_d$ the drag coefficient. 
The rate of change of energy per unit area due to skin drag is  proportional to $|u_{\text{orb}}-v|^3 $ \citep{shen1998wave}, where  $u_{\text{orb}}=U \sin{(\phi)}$ is the wave orbital velocity at the ice-water interface, here evaluated at the free surface under the hypothesis of thin sea ice, and $v$ is the ice drift velocity.
The magnitude of the orbital velocity is  $U =a \Omega$, where $\Omega=gk/\omega $, and $ \phi $ is time-dependent phase.
Phase averaging, we obtain:
\begin{equation}\label{skindrag}
s_{\text{sd}} = - \frac{ C_{d} }{g} \frac{1}{2\pi} \int_0^{2\pi} \left| a\Omega\sin(\phi) -v \right|^3 d\phi.
\end{equation}

We consider a thin semi-infinite homogenised sea ice layer drifting at a constant velocity $v$ and arbitrarily set a frame of reference moving at $v$, i.e.~$\hat{x}=x-vt$, with $\hat{x}$ the distance from the edge of the sea ice. 
Taking the steady state system in the moving frame with a constant wave forcing at the ice edge $a(0)=a_0$, the one-dimensional wave energy transport equation Eq.~\ref{energyequationa} in the ice region ($\hat{x}\ge0$) reduces to:

\begin{equation}\label{aeq}
 \frac{da}{d\hat{x}} = - \alpha \, a-  \frac{\Gamma}{a} \int_0^{2\pi} \left| a \Omega \sin(\phi) -v \right|^3 d\phi \,  , 
\end{equation} 
with
\begin{equation}
\Gamma = \frac{C_d}{2\pi g (c_g-v)} \, ,
\quad \text{and} \quad
\alpha=\frac{c_g}{c_g-v} \alpha_\textrm{exp},
\end{equation} 
being the attenuation, therefore remaining exponential in the moving frame.

\subsection{Solving the Amplitude Profile}
Equation~\ref{aeq} represents a challenging nonlinear ordinary differential equation.
Here we aim to derive an analytical solution and to facilitate the mathematical treatment we assume the parameters $ C_d $ and $ \alpha $ to be spatially homogeneous (constant).
The model can be extended for heterogeneous conditions but numerical solutions should be sought. 
A polynomial in $ a $ for the integral of the skin drag  (Eq.~\ref{skindrag})  can be found explicitly in closed form for $|v| \ge a\Omega $, see Appendix \ref{appA}.
In effort to enable closed form analytical solutions  we seek to an approximate polynomial form using an asymptotic expansion of the integrand for  $|v |\ll a\Omega $ that are then extended into the region $|v|< a\Omega $  (see Appendix \ref{appA}). We obtain: 
\begin{equation}\label{integral}
I = \int_0^{2\pi} | a\Omega \sin(\phi) -v|^3 d\phi 
\begin{cases}
      \approx \frac{8}{3} (a\Omega)^3 + 12 a\Omega v^2 , & \text{if }  |v |< a\Omega;  \\
      = 3\pi (a\Omega)^2 |v| + 2\pi |v|^3 , & \text{if } |v| \ge a\Omega .
\end{cases}
\end{equation}

Note that, in oceanic conditions, $c_g$ is two orders of magnitude lager than the wind and current induced drift, i.e. $\mathcal{O}(10)\,\mathrm{m/s}$ versus $\mathcal{O}(0.1)\,\mathrm{m/s}$, therefore $|v|\ll c_g$ implying that, with $\alpha \text{ and }\Gamma $  positive, Eq.~\ref{integral} is strictly positive, therefore the amplitude $ a $ decreases monotonically in Eq.~\ref{aeq}.
For $a_0 \Omega >|v|$, we define $\hat{x}^*$ to be the location into the ice domain such that $a(\hat{x}^*) \Omega = |v|$. 
Therefore, the condition $|v| \lessgtr a\Omega $ corresponds to $\hat{x} \lessgtr \hat{x}^*$ in the physical domain. In contrast, for $a_0\Omega\le|v|$, one obtains $\hat{x}^*=0$. 

Solving Eq.~\ref{aeq} with the integral approximation Eq.~\ref{integral}, the amplitude solutions are written in piecewise form as: 
\begin{equation} \label{eq:AB}
a(\hat{x}) \approx a_A(\hat{x}) \, \mathbf{1}_{\{\hat{x} < \hat{x}^*\}} + a_B(\hat{x}) \, \mathbf{1}_{\{\hat{x} \ge \hat{x}^*\}},
\end{equation}
where $\mathbf{1}$ is the indicator function, equal to $1$ if the condition is true and $0$ otherwise. The piecewise solution components result in: 
\begin{equation}\label{firstbranch}
a_{A}(\hat{x}) =
\begin{cases}
\displaystyle \frac{3\sqrt{\Delta}}{16\Gamma \Omega^3} 
\tan \Biggl(-\frac{\sqrt{\Delta}}{2} \hat{x} 
+ \arctan\Bigl(\frac{16\Gamma \Omega^3 a_0 + 3\alpha}{3\sqrt{\Delta}}\Bigr)\Biggr) 
- \frac{3\alpha}{16\Gamma \Omega^3}, & \delta > 1, \\[1.3ex]
\displaystyle \frac{32 \Gamma \Omega^3 a_0 - \left(16 \alpha \Gamma \Omega^3 a_0 + 3 \alpha^2\right)\hat{x}}
      {32 \Gamma \Omega^3 + \left(32 \Gamma \Omega^3 a_0 + 6 \alpha \right)\hat{x}}, & \delta = 1, \\[1.3ex]
\displaystyle \frac{r_-(a_0-r_+)\exp\left(\frac{8 \Gamma \Omega^3}{3}(r_- -r_+) \hat{x} \right) - r_+(a_0-r_-)}
        {(a_0-r_+)\exp\left(\frac{8 \Gamma\Omega^3}{3}(r_- -r_+) \hat{x} \right) - (a_0-r_-)},  & \delta < 1,
\end{cases}
\end{equation}
\begin{equation}
a_B(\hat{x}) = \left[ e^{-(6 \pi \Omega^2 |v|\Gamma + 2\alpha)(\hat{x}-\hat{x}^*)}\left(\frac{v^2}{\Omega^2}+ \frac{2\pi |v|^3 \Gamma}{3\pi \Omega^2 |v| \Gamma + \alpha}\right) - \frac{2\pi |v|^3 \Gamma}{3\pi \Omega^2 |v| \Gamma + \alpha} \right]^{\tfrac{1}{2}} \, .
\end{equation}
Here  $\Delta \equiv \alpha^2 (\delta^2-1)$, $r_\pm \equiv -3(\alpha \pm \sqrt{-\Delta})/(16 \Gamma \Omega^3)$, and
\begin{equation}
    \delta =  \frac{8\sqrt{2} \Gamma \Omega^2 |v|}{\alpha}\,,
\end{equation}
which controls the roots of the quadratic polynomial in $ a $ affecting the solution in the $ \hat{x} < \hat{x}^* $ region (see Appendix \ref{appB}).
Note that the physical interpretation of $ \delta $ will be discussed in \S2.2.

The solution has a few interesting features.
The amplitude profile is continuous in $\hat{x}$ (see Appendix \ref{appB}). {The solution is nonlinear, i.e. drag effects and other mechanisms cannot be added as linear superposition.
For $\hat{x}< \hat{x}^*$, the different branches of the solution in Eq.~\ref{firstbranch}  differ in form but possess similar qualitative behaviour, where we note that $\hat{x}^*$ depends on $\delta$. 
For a wave with initial orbital velocity magnitude $U_0 <|v|$ we obtain the solution in region B, $a_B$,  applies to the entire domain $\hat{x}\ge 0$ .

A distinctive consequence of such solution is the existence of an extinction location
\begin{equation}
\hat{x}_{\text{end}} = \hat{x}^* + \frac{8}{3 \sqrt{2} \pi \alpha \delta +8} \log\left(\frac{5}{2} + \frac{4 \sqrt{2}}{\pi \delta}\right)\,,
\end{equation}
such that $a_B(\hat{x}=\hat{x}_{\text{end}})=0$.The extinction location is always in region B because it is where amplitude tends to zero and the condition $a\Omega<|v|$ is satisfied. 
The extinction location is unique to dissipation due to drift-induced skin drag and defines the extent of the wave affected region.
It moves closer to the sea ice edge for increasing $C_D$ coefficient and  $\alpha$ (see solution in red, for $v=0.05$\,m/s and $\alpha=7\times10^{-6}{ \text{ m}^{-1}}$ \, [$\delta=3.4$], vs in blue, for $v=0.05$\,m/s and $\alpha=0$ [$\delta=\infty$], in Fig.~\ref{fig:figure1}a, while its dependence on drift is not trivial. Such extinction locations are not present in the solutions of \citet{kohout2011wave} and \citet{herman2019wave}, shown dashed green and orange respectively in Fig.~\ref{fig:figure1}a), for which $\alpha=0$ and $v=0$.

In the special case of no drift, i.e. $v=0$, the solution reduces to: 
\begin{equation}\label{zerodrift}
a(\hat{x}) = \frac{3a_0 \alpha e^{-\alpha \hat{x}}}{3\alpha+8a_0\Gamma\Omega^3 -8a_0\Gamma\Omega^3 e^{-\alpha \hat{x}}}.
\end{equation}
Taking $C_d=0$, i.e.\,no drag, we recover the exponential attenuation at a rate $\alpha$ whereas for  $\alpha \rightarrow 0$ the solution converges to the one of \citet{herman2019wave}.
Therefore, Eq.~\ref{zerodrift} is an extension \citet{herman2019wave} that accounts for additional mechanisms.
As a result, our solution predicts lower amplitudes compared to \citet{herman2019wave}, see Fig.~\ref{fig:figure1}a.

\begin{figure}
    \centering
    \includegraphics[width=1\textwidth]{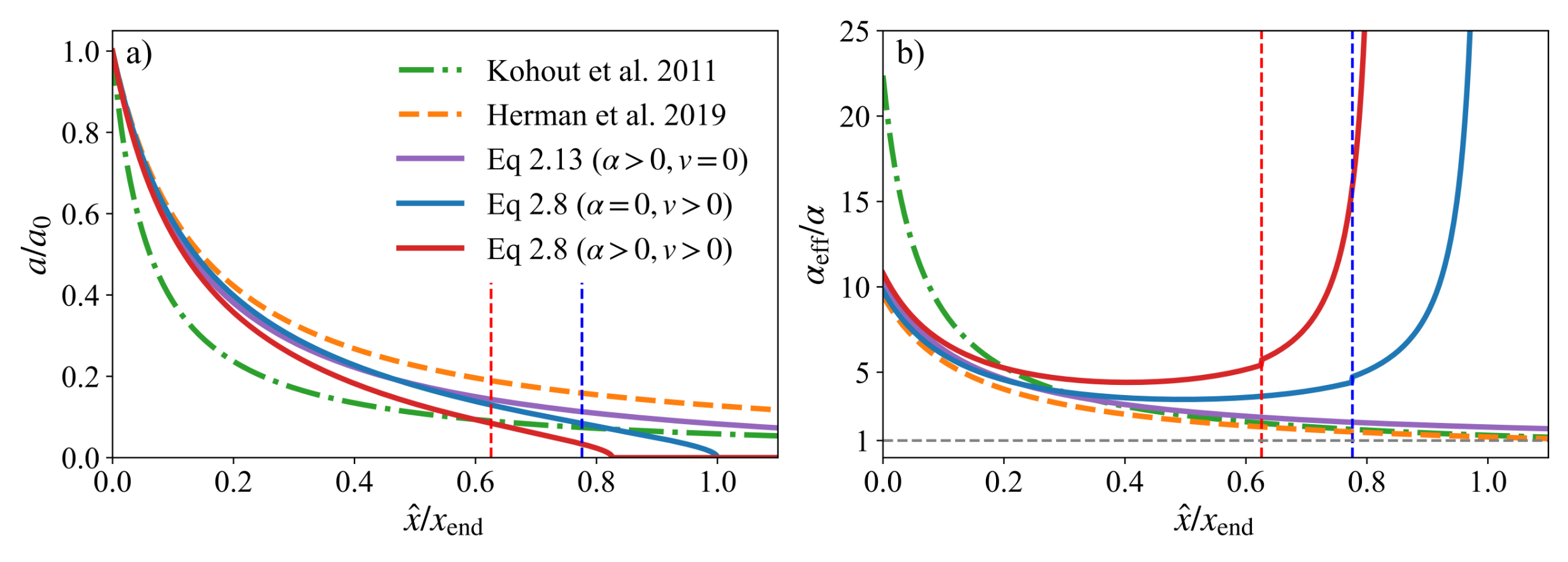}
\caption{Sample amplitude (a) and corresponding attenuation (b) profiles according to our model and formulations by \citet{kohout2011wave} and \citet{herman2019wave}.
Dashed lines denote $x^*$ which denotes the separation between region A and B. Amplitude is normalised by the initial amplitude and distance by $x_\text{end}$ for $\alpha=0$. Attenuation rate is normalised by $\alpha$. Solutions are shown for $a_0=1$\,m, $C_d=0.05$, $v=0.053\,\mathrm{m\, s^{-1}}$, $\omega = 0.63\,\mathrm{s^{-1}}$ 
, $\alpha=7.2\times10^{-6} \mathrm{\,m^{-1}}$.}  \label{fig:figure1}
\end{figure}

Fig.\,\ref{fig:figure1}a displays the normalised amplitude as a function of the distance into the MIZ normalised by the value $ x_{\text{end}} $ computed for the non zero drift and exponential attenuation case ($v=0.053 \mathrm{m\, s^{-1}}$ and $\alpha=0 $; the case displayed in blue). 
The attenuation is stronger when drift is present   as shown in red ($v>0$ and $\alpha=7\times10^{-6} $) compared to the solution in purple with ($v=0$\,m/s and $\alpha=7\times10^{-6}$).
The increased attenuation due to other mechanisms also manifests when drift is present, see solutions in blue vs in red.

\subsection{Predicting the Attenuation Rate}

To allow better comparison with attenuation results for non-drifting sea ice as they appear in the literature \citep{kohout2011wave,herman2019wave} and better highlight the attenuation characteristics,

we compute the effective amplitude attenuation rate in the moving frame in the ice covered region defined as
\begin{equation}
\alpha_\text{eff} \equiv -\frac{1}{a} \frac{da}{d \hat{x}}=   \alpha \, +  \frac{\Gamma}{a^2} I  \,   .
\end{equation}
Using the integral estimation (Eq.~\ref{integral}) we obtain: 
\begin{equation}\label{attenbranch}
\alpha_\text{eff}(\hat{x}) 
\begin{cases}
     \approx \alpha+\frac{8}{3} a \Omega^3  \Gamma + \frac{12  \Omega v_{}^2  \Gamma}{a}    , & \text{if }  \hat{x}<\hat{x}^*;  \\
      = \alpha+3\pi  \Omega^2 |v| \Gamma + \frac{2\pi |v|^3  \Gamma}{a^2}   , & \text{if } \hat{x}\geq \hat{x}^*.
\end{cases}
\end{equation}

Note that, by extending asymptotic solution of $I$ for $|v|<a\Omega $ to $|v|=a\Omega $, the decay rate inherits a jump discontinuity at $\hat{x}^*$, shown in Fig.~\ref{fig:figure1}b, with the suitability of the integral approximation justified by discontinuity magnitude being only 7\%.
The nonlinearity between skin drag and other exponential decay mechanisms leads to the lower bound
\begin{equation}
\alpha_\text{eff} \geq \alpha +8 \sqrt{2} \Gamma \Omega^2 |v| = \alpha (1+\delta) \,,
\end{equation}

with the global minimum located in the region $\hat{x}<\hat{x}^*$ if $ a_0\Omega \geq 3\sqrt{2} |v|/2$, whereas for lower orbital velocity at the sea ice edge the attenuation monotonically increases.
This result allows for a physical interpretation of the dimensionless parameter $ \delta $: for $\delta >1$ wave attenuation is dominated by drift-induced skin drag and for $\delta <1$ all other mechanisms dominate.

For drifting sea ice, the attenuation rate grows unbounded in the neighbourhood of the extinction location (see solutions in red and blue in Fig.~\ref{fig:figure1}b), i.e.~ $\alpha_\text{eff}(\hat{x}) \sim [2(x_{\text{end}} - \hat{x})]^{-1}$, therefore the extinction location affects the magnitude of the attenuation rate in this region.
When drift is absent, $\hat{x}^* \rightarrow \infty$ and the solution for $\hat{x}<\hat{x}^*$ applies to the entire domain. 
As a consequence of our model, skin drag dominates attenuation close to the sea ice edge whereas other processes dominate deeper into the sea ice, where we retrieve the classical exponential attenuation, i.e. $\alpha_\text{eff} \rightarrow \alpha$ for $\hat{x} \rightarrow \infty$ (see solution in purple in Fig.~\ref{fig:figure1}b; the grey dashed line denotes $\alpha_\text{eff} = \alpha$).

\section{Comparisons to Antarctic Measurements}

In this section comparisons are made with measured amplitudes and derived attenuations in the Antarctic sea ice \citep{brouwer2022altimetric,voermans2025finely}. 
The observations comprises quality controlled ICESat-2 satellite transects of wave height (both the total wave heights and the spectral wave heights corresponding to frequency bands at periods $T=9,12,15,18 \text{ s}$) across the MIZ throughout 2019.
Note that transects seldom start at the ice edge (possibly because they fail quality control in intermediate sea ice concentrations) and, in computing attenuation rates, transects are assumed to align with the wave direction \citep{voermans2025finely}.
Drift is not measured by ICESat-2.

To compare observations to model results, which are derived for a monochromatic wave, we attribute all energy to the most energetic frequency component measured over the transect, normalising amplitude and attenuation rate by the first available measurement from the ice edge,  $a_V^1$ and $\alpha^1_V$. The ice edge is taken to coincide be the start of each transect. To allow for a straightforward comparison of the model with the observations we use $x$ in place of $\hat{x}$ noting that their difference is negligible for small drift velocities. Moreover, in agreement with the observations, transects are normalised by MIZ extent $x_{\text{MIZ}}$. In this regard is important to note that MIZ extent is reported independently from the wave amplitude, i.e. low amplitude readings were excluded due to quality control \citep{voermans2025finely}.

Drift and drag are used as fitting parameters, within the range of available Antarctic sea ice observations, i.e.~$v=\mathcal{O}(10^{-1})\, \text{ m/s }$ and $C_d = \mathcal{O}(10^{-2})$, together with $\alpha$. While least squares optimisations was attempted, the noise in the observational data prevented reasonable fitting throughout the transect. For illustrative purposes,  we performed manual tuning for each transect of the parameters within ranges of characteristic values for the MIZ. The drift is assumed to be constant along the transect, justified by slow spatial variation over 10--100\,km in Antarctic sea ice \citep{holland2012wind}.
The water density is set $\rho=1025 \text{ kg/m}^3$ and deep water dispersion relation used, namely $ \Omega= \omega $.

\subsection{Individual Transects}

The two transects reported in \citet{voermans2025finely} are analysed in detail.
Transect I, taken on 26 December 2019 at $85^\circ \mathrm{W},\; 69^\circ \mathrm{S}$, is 61.5\,km long and waves have peak period 15\,s with amplitude $a^{1}_V = 0.45  \text{ m and attenuation},\alpha^1_{\mathrm{V}} = 1.42 \times 10^{-5} \, \mathrm{m}^{-1}$.
Transect II, taken on 24 May 2019 at $88^\circ \mathrm{W},\; 68^\circ \mathrm{S}$, is longer ($x_{\text{MIZ}}=137.5 \text{ km}$) and waves have a shorter peak period ($T=$12\,s) with amplitude $a^{1}_V = 0.55  \text{ m and attenuation},\alpha^1_{\mathrm{V}} = 2.43 \times 10^{-6} \, \mathrm{m}^{-1}$.

For Transect I the parameters $ v=0.26 \text{ m/s}, C_d = 6.0 \times 10^{-3}$ and $\alpha = 5.0\times 10^{-6}$ provide good qualitative and quantitative agreement in terms of amplitude decay (see Fig.~\ref{fig:figure2}a). Similar performances are achieved by the the non-drifting model, i.e. $v=0$, and the exponential decay.
However, the model with drift outperforms the others reproducing the observed increase in effective attenuation rate along the transect (between $  x/x_{\text{MIZ}} \approx0.2 \text{ and }  0.6$; see Fig.~\ref{fig:figure2}b).  
Note that for the chosen parameters, drift is stronger than the orbital velocity throughout the transect ($\lvert v \rvert / U_0 \approx 1.2$), i.e. only the branch $x>\hat{x}^*$ of Eq.~\ref{attenbranch} is used.
Remarkably, the model predicts an extinction location consistent with the observed wave-affected sea ice extend, i.e.\,at $x/x_{\text{MIZ}} \approx1$.

\begin{figure}
    \centering
    \includegraphics[width=1\textwidth]{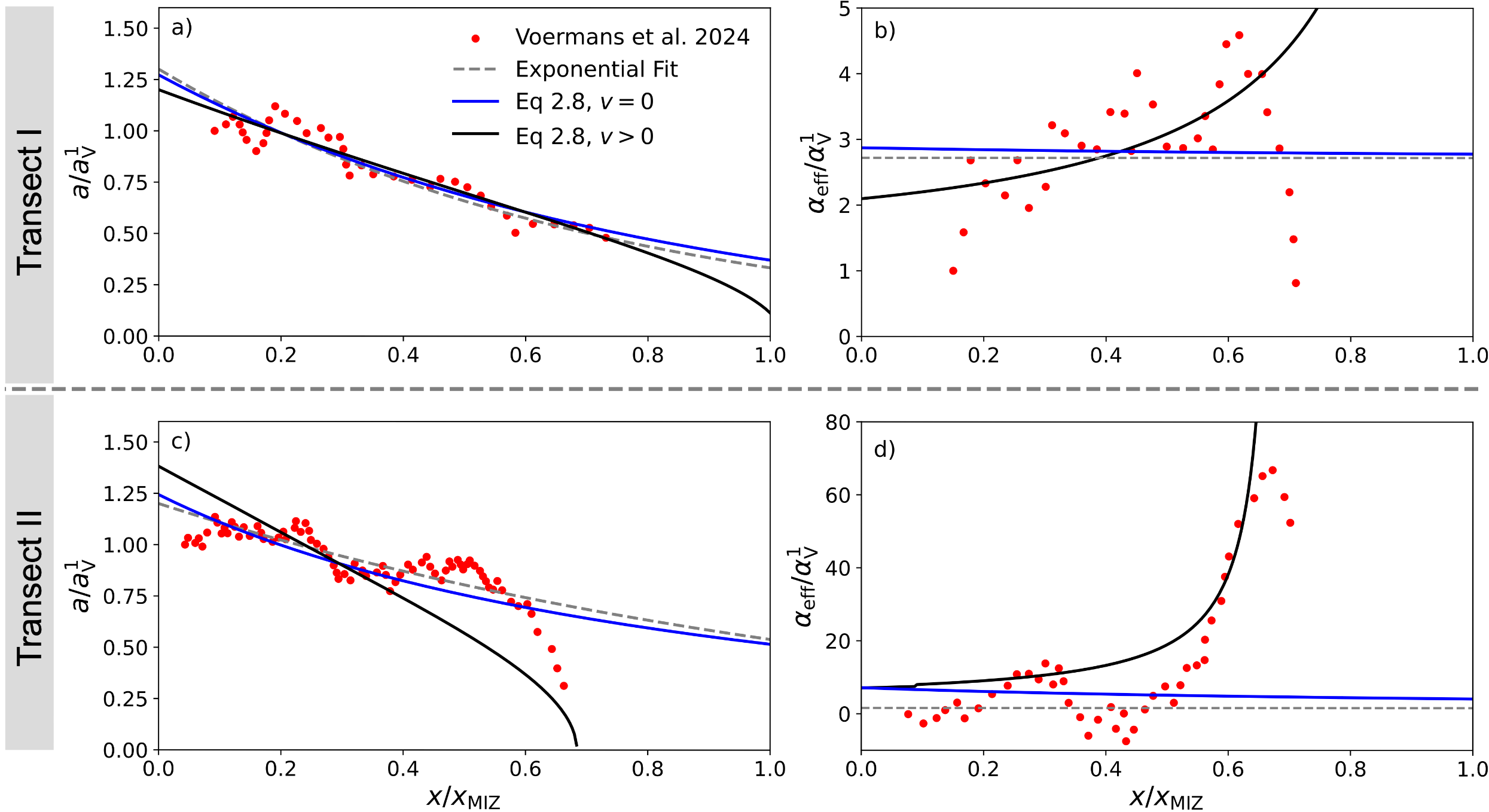}
    \caption{Amplitude (left) and corresponding attenuation profiles  (right) for Transect A (top) and B (bottom). Measurements (red dots) are shown against model predictions. Distance is normalised with respect to the wave-affected sea ice extent $x_{\text{MIZ}}$ and amplitude/attenuation with respect to the measurement closest to the sea ice edge as reported in \citet{voermans2025finely}.}
    \label{fig:figure2}
\end{figure}

For Transect II the parameters $v=0.49 \text{ m/s}, C_d = 2.8 \times 10^{-3}$ and $\alpha = 1.0\times 10^{-7}$ qualitatively reproduce the measured mild amplitude decay up to $ x/x_{\text{MIZ}}<0.6$ followed by a sharp decline (see Fig.~\ref{fig:figure2}c).
Other models, i.e. the exponential and non-drifting model better reproduce the attenuation near to the ice edge, but fail to capture the sharp decline deeper into the sea ice. 
The predictive performance of the drifting model is clearly highlighted by the attenuation rate, shown in Fig.~\ref{fig:figure2}d, where steady attenuation rate near the ice edge followed by rapid growth for $x/x_{\text{MIZ}} >0.6$ are well captured qualitatively and quantitatively.
Note that the observed attenuation rate fluctuates resulting in negative values close to the edge and again at $x/x_{\text{MIZ}}\approx 0.4$, and models can not capture this behaviour.
A jump discontinuity appears at $x/x_{\text{MIZ}} \approx0.1$, consistent with the choice of $v$ resulting in $\lvert v \rvert / U_0 \approx 0.9$ and the presence of $\hat{x}^*$ within the domain.
Unlike our formulation, all other models predict either a decreasing attenuation rate (no drift model) or constant attenuation (exponential model), as shown in Fig.~\ref{fig:figure2}d.

In summary, Transect I and II showcase two different regimes, in Transect I drag attenuation is weak ($\delta=0.6$) whereas in Transect II it is the dominant mechanism ($\delta=8.0$).
Therefore, as expected, in Transect I deviations from exponential attenuation are weaker than in Transect II however, also in Transect II, inclusion of exponential attenuation through $\alpha$ provides significantly improved predictions.
When drag dominates, the extinction location $x_{\text{end}}$ occurs earlier and the spike in attenuation is more prominent.

Our model is developed for homogeneous sea ice conditions, implying constant $C_d$ along the transect. While this hypothesis is more likely to hold in the winter season 
(Transect II) its applicability to Transect I is more debatable. Moreover, direct comparison with individual observations is made difficult by inherent measurement uncertainty, including misalignenment of the transect with the wave direction \cite{voermans2025finely}, lack of information on the local sea ice properties, and possible presence of other forcing mechanisms in the wave transport equation, e.g. wind forcing. Nevertheless, we highlight that our simple model captures the observed feature of wave attenuation in the MIZ.

\subsection{Averaged Attenuation}

\citet{voermans2025finely} report average attenuation profiles across seasons and sectors of the Antarctic MIZ.
To reproduce the observed variability, we drive our model, i.e.\,Eq.~\ref{attenbranch}, with $N_0=3\times10^5$ random realisations with Gaussian distributed drift velocity (with mean $\mu_v$ and standard deviation $\sigma_v$), constant throughout the transect. Each realisation corresponds to an individual transect and $N$ denotes the number of transects at any given location according to our model, the decrease is due to individual transects reaching the extinction location.
The simulations replicate the variability of drift around Antarctica \citep{kwok2017sea}.
The attenuation rate is computed for the wave component $T = 12\ \mathrm{s}$ (angular frequency $\Omega = 0.52 \,\mathrm{s}^{-1}$ and wavelength $\lambda = 225$\,m) and we set the wave amplitude at the sea ice edge to $a_0=1 \text{m}$. The other parameters are $C_d = 0.02$ and $\alpha = 7.0 \times 10^{-6}\ \mathrm{m}^{-1}$,  to match the averaged measured attenuation rate at the sea ice edge. While the drag coefficient is higher than the one chosen for the individual transects in the previous section, its value is within the range of observed values for the MIZ  \citep{mcphee1979effect}. Moreover, the chosen value allows to quantitatively match the observed attenuation rate close to the sea ice edge as reported in \citet{voermans2025finely}. Smaller values of drag would give rise to a similar qualitative behaviour.  For the comparison, only $a\geq 0.05 \text{m}$ are averaged, in  accordance with the methodology used by \citet{voermans2025finely} to exclude observations with high uncertainty.

For drift $\mu_v=0.22$\,m/s ($\mu_v/U_0=0.42$) and standard deviation $\sigma_v=0.03$\,m/s (Fig.~\ref{fig:figure3}a) the attenuation only slightly increases up to $x/\lambda \approx 550$, consistent with the fact that simulations are in weak drag regime ($\delta=0.42$).
From $x/\lambda \approx 550$ to $x/\lambda \approx 700$ the increase is more noticeable, up to $\approx3$ times the attenuation rate at the sea ice edge, in qualitative agreement with simulations for Transect I (cf. Fig.~\ref{fig:figure2}b), before stabilising farther in the sea ice domain. 
The spread of individual profiles, shaded in grey, also grows when attenuation rate increases.
Note that the increase in attenuation corresponds to a reduction in available measurements of amplitude over threshold (shown in red) and, similarly, individual amplitude profiles reaching the extinction location (distribution shaded in blue).  

\begin{figure}
    \centering
    \includegraphics[width=1\textwidth]{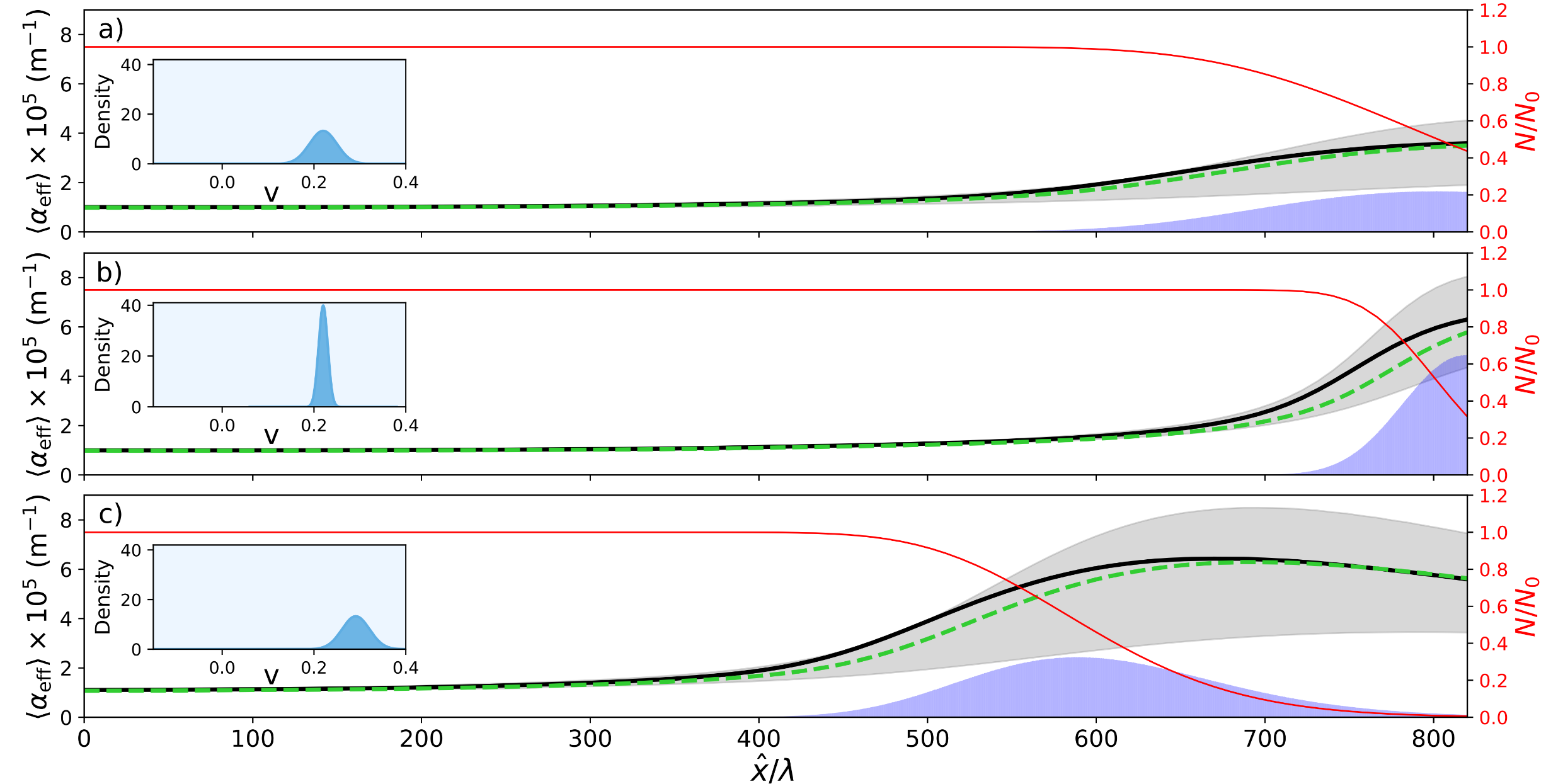}
    \caption{Averaged attenuation rates (in black; for negative velocity in green) and interquartile range (shaded in grey). On the right axis, number of measurements (in red) and distribution of the extinction location (shaded in blue; not to scale).
    Insets show the corresponding drift velocity distribution: (a) $\mu_v=0.22$\, m s$^{-1}$ $\sigma_v=0.03$\, m s$^{-1}$; (b) $\mu_v=0.22$\, m s$^{-1}$ $\sigma_v=0.01$\, m s$^{-1}$; (c) $\mu_v=0.29$\, m s$^{-1}$ $\sigma_v=0.03$\, m s$^{-1}$. }
    \label{fig:figure3}
\end{figure}

With the same average drift ($\mu_v=0.22$\,m/s) but lower variability (from $\sigma_v=0.03$\,m/s to $\sigma_v=0.01$\,m/s; Fig.~\ref{fig:figure3}b) the same qualitative behaviour is observed. 
The increase in attenuation is sharper, i.e. it starts deeper in the sea ice (at $\approx650$ wavelengths) and reaches higher values ($\alpha_{\textrm{eff}}\approx5.5 \times 10^{-4}$ vs $\alpha_{\textrm{eff}}\approx3.5 \times 10^{-4}$).

The increase in drift ($\mu_v=0.29$\,m/s, $\mu_v/U_0=0.55$  and $\sigma_v=0.03$\,m/s; Fig.~\ref{fig:figure3}c), results in higher $\delta=0.55$, still in weak drag regime, and moves the increase in attenuation rate closer to the sea ice edge, from $x/\lambda \approx 450$ to $x/\lambda \approx 550$. A slightly higher attenuation is also achieved compared to the previous cases, up to $\alpha_{\textrm{eff}}\approx6\times10^{-4}$, after which decreases slightly. For smaller drag coefficients results the increase in attenuation occurs further into the ice field.
The number of observations decreases sharply due to individual transects reaching their extinction location (cf. Fig.~\ref{fig:figure1}b), and only less than 0.1\% remain after 800 wavelengths, also explaining the larger variance in effective attenuation.

In Fig.~\ref{fig:figure3} the mean attenuation rates for negative mean velocity are also reported in dashed green line.
The direction of the drift has only marginal effect on the effective attenuation profile, i.e.\,the growth of attenuation rate is deeper into the sea ice (by $\approx10$ wavelengths) but does not affect its magnitude, with similar effect on available measurements and extinction location distributions.  
This is expected for oceanic conditions, as $\alpha\approx \alpha_\textrm{exp}$ and $\Gamma \approx C_d/( g c_g 2\pi )$, hence the dependence of direction of the drift is negligible for operational purposes. 

Fig.~\ref{fig:figure3}c is remarkably similar to the measured averaged attenuation in \citet{voermans2025finely}, i.e. a linear increase followed by slight decline, also matching the range of observed attenuation rates.
Unlike in the measurements, in the model all individual transects are available from the edge but the decrease of available observations in \citet{voermans2025finely} deeper into the sea ice is also captured.
It can be argued that the presence of drift limits the extent of the wave-affected area, and its extent in our model ($\approx185$\,km; Fig.~\ref{fig:figure3}c) is consistent to the MIZ width of the order of 200\,km as reportedin observations \citep{brouwer2022altimetric} and model simulations \citep{day2024jgr}.

\section{Conclusions}

We derived a mechanistic model for wave attenuation that accounts for both phenomenological exponential attenuation mechanisms and dissipation due to drag at the ocean-sea ice interface, as well as the effect of sea ice drift, therefore combining and extending previous models in which drift was excluded \citep[cf.][]{shen1998wave,kohout2011wave,herman2019wave}.
We identified the parameter $\delta$ which governs attenuation dominated by drift-induced drag mechanisms ($\delta>1$) and exponential mechanisms ($\delta<1$).
Analytical solutions reveal that in drift-induced drag dominated regime, the amplitude eventually dies off, unlike prediction based on exponential attenuation, and the attenuation profile spikes close to the extinction location.

The model agrees well, both qualitatively and quantitatively with pan-Antarctic measurements of attenuation rate recently reported by \citet{voermans2025finely} when sea ice is allowed to drift slowly ($\mu_v/U_0=0.55$), even for small $\delta$, i.e.~$\delta=0.55$.
In particular, drift is responsible for the increase in wave attenuation rate deeper into the sea ice and, to a certain extent, regulates the {width of the wave-affected sea ice} which is in quantitative agreement to the MIZ extent (100--200\,km) reported for observations \citep{brouwer2022altimetric} and model simulations \citep{day2024jgr}.

The quantitative agreement against individual transects is reasonable with limitations.
Differences with the measured observations can be attributed to experimental uncertainty  and current model assumptions, i.e.~no interaction between wave components, no presence of other forcing mechanisms, homogeneous drift and sea ice properties along the transect. Some of the limitations of the model could be sought by seeking numerical solutions for the governing equation (Eq.~\ref{aeq}) and allowing for spatially heterogenous parameters. In this regard, our modelling framework allows for easy implementation into operational models and the biggest limitation, common to many wave attenuation models, is the lack of physically-informed sea ice parameters along the MIZ.

Ultimately the model provides a solid theoretical foundation to better interpret observations of wave attenuation in drifting sea ice as found around Antarctica, and prompts new concurrent measurements of waves, drift, and sea ice properties for further validation.
Moreover, while only drag has been studied explicitly, the approach can be extended to better specify other wave attenuation mechanisms for integration in emerging coupled ocean-wave-sea ice models as the ones currently being developed in most weather forecasting agencies \citep
{kousal2025bams}, e.g. ECMWF.

\appendix

\section{Integral Approximation} \label{appA}

The integral $I$ in Eq.\ref{integral}:}
\begin{equation}
I = \int_0^{2\pi} |U \sin\phi - v|^3 \, d\phi,
\end{equation}
is treated differently for $|v|\ll a\Omega$ (or equivalently $|v|\ll U$) and $|v| > a\Omega$ (or equivalently $|v|> U$).

\subsection{Case $ |v|\ll U$.}

We expand the integrand as:
\begin{equation}
|U \sin\phi - v|^3  =  |U \sin\phi|^3 - 3v U^2 \sin^2\phi \,\mathrm{sgn}(\sin\phi)\sin\phi + 3 v^2 U |\sin\phi| + O(v^3).
\end{equation}

The second term has zero contribution towards $I$. Therefore, we approximate the integral as a polynomial in $U$:
\begin{equation}
I \approx \int_0^{2\pi} \left( U^3 |\sin\phi|^3 + 3 U v^2 |\sin\phi| \right) d\phi = \dfrac{8}{3} U^3 + 12 U v^2
\end{equation}
While strictly valid only for $|v|\ll U$ we use this approximation for $|v|\le U$.

\subsection{Case: $ |v|> U $}
The integrand only depends on the sign of $v$. Therefore it can be written as:
\begin{equation}
|U \sin\phi - v|^3 = \text{sign}(v) \, (v-  U \sin\phi )^3.
\end{equation}
Note that some of the terms have zero contribution towards $I$ due to symmetry. We write the integral as a polynomial:
\begin{equation}
I =\int_0^{2\pi} \left( |v|^3  + 3 |v| U^2 \sin^2\phi \right) d\phi
\end{equation}
\begin{equation}
I = 2\pi |v|^3 + 3\pi U^2 |v|
\end{equation}

\section{Analytical Solutions} \label{appB}

The integral approximation $I$ is piecewise, i.e. depends on the relationship between $U$ and $v$.
As a result, different governing equations emerge for solutions in region A, corresponding to $a\Omega \ge |v|$ , and region B for $a\Omega < |v|$.
We first solve the governing equation for $a$ in region A and then impose continuity on the amplitude to obtain the solution in region B.

\subsection{Region A.}

In Region A, defined as the region in which orbital velocity is greater than the drift ($a\Omega \ge |v|$), if it exists we impose the initial condition $a(\hat{x}=0)=a_0$. 

The governing equation in its differential and integral form is: 
\begin{equation}
\frac{da_A}{d\hat{x}}= -\alpha a_A-\tfrac{8}{3}\Gamma \Omega^3a_A^2 - 12\Gamma \Omega |v|^2 \equiv -f(a),
\end{equation}
\begin{equation}
\int_{a_0} ^{a_A} \frac{d a'}{\tfrac{8}{3}\Gamma \Omega^3a'^2+\alpha a'+12\Gamma \Omega |v|^2} = -\hat{x}.
\end{equation}
The sign of the discriminant of $f$, denoted as $\Delta$, sets the form of solution.

The roots of $f$ are $r_{\pm}$.

The solutions are categorised according to $\Delta$:
\begin{equation}
a_{A}(\hat{x}) =
\begin{cases}
\displaystyle \frac{3\sqrt{\Delta}}{16\Gamma \Omega^3} \tan \Biggl(-\frac{\sqrt{\Delta}}{2} \hat{x} + \arctan\Bigl(\frac{16\Gamma \Omega^3 a_0 + 3\alpha}{3\sqrt{\Delta}}\Bigr)\Biggr) - \frac{3\alpha}{16\Gamma \Omega^3}, & \Delta > 0, \\[1.5ex]
\displaystyle \frac{32 \Gamma \Omega^3 a_0 - \left(16 \alpha \Gamma \Omega^3 a_0 + 3 \alpha^2\right)\hat{x}}
      {32 \Gamma \Omega^3 + \left(32 \Gamma \Omega^3 a_0 + 6 \alpha \right)\hat{x}}, & \Delta = 0, \\[1.5ex]
\displaystyle \frac{r_-(a_0-r_+)\exp\left(\frac{8 \Gamma \Omega^3}{3}(r_- -r_+) \hat{x} \right) - r_+(a_0-r_-)}
        {(a_0-r_+)\exp\left(\frac{8 \Gamma\Omega^3}{3}(r_- -r_+) \hat{x} \right) - (a_0-r_-)},  & \Delta < 0.
\end{cases}
\end{equation}
Because the relationship between $\Delta$ and $\delta$, we can rewrite the condition as function of $\delta$ as in Eq.~\ref{firstbranch}.

We define $\hat{x}^*$ as the location where $U(\hat{x}^*) = |v|$, or equivalently $a(\hat{x}^*)\Omega = |v|$. Therefore:
\begin{equation}
    \hat{x}^* = \max\!\left\{\, a^{-1}\!\left(\frac{|v|}{\Omega}\right),\, 0 \,\right\},
\end{equation}
denotes the extent of region A.

\subsection{Region B}

In region B we enforce continuity of the amplitude at $\hat{x}^*$. Note that if the orbital velocity at the ice edge is lower that the drift only region B exists. Using the integral approximation, the differential and integral form of the governing equation are:
\begin{equation}
\frac{da_B}{d\hat{x}} \;=\; -\left(\alpha + 3\pi \Gamma \Omega^2 |v|\right)a_B \;-\; \frac{2\pi \Gamma |v|^3}{a_B},
\end{equation}
\begin{equation}
\int_{a_A(x^*)}^{a_B} \frac{a'}{\left(\alpha + 3\pi \Gamma \Omega^2 |v|\right)a'^2 + 2\pi \Gamma |v|^3}\, da' \;=\; x^*-\hat{x}.
\end{equation}

The integrand yields:
\begin{equation}
a_B(\hat{x}) = \left[ e^{-(6 \pi \Omega^2 |v|\Gamma + 2\alpha)(\hat{x}-\hat{x}^*)}\left(\frac{v^2}{\Omega^2}+ \frac{2\pi |v|^3 \Gamma}{3\pi \Omega^2 |v| \Gamma + \alpha}\right) - \frac{2\pi |v|^3 \Gamma}{3\pi \Omega^2 |v| \Gamma + \alpha} \right]^{\tfrac{1}{2}}, \qquad \hat{x}\geq \hat{x}^*.
\end{equation}

\backsection[Acknowledgements]{This work was supported by the Natural Environment Research Council and the ARIES Doctoral Training Partnership [grant number NE/S007334/1]. 
RR and AA acknowledge funding from Royal Society (IEC\textbackslash R3\textbackslash 243016).
This work was stimulated by discussions during the \textit{``Maths of Sea Ice''} meeting, made possible by the Isaac Newton Institute for Mathematical Sciences.
DP is supported by EPSRC Grant No. EP/Y021118/1 and by the ExtreMe Matter Institute EMMI at the GSI Helmholtzzentrum fuer Schwerionenphysik, Darmstadt, Germany.}

\backsection[Declaration of interests]{The authors report no conflict of interest.}

\end{document}